\documentclass[pre,aps,twocolumn,superscriptaddress,amsmath,showpacs,floatfix]{revtex4}
\usepackage{graphicx}
\begin{document}
\title{Muscle contraction and the elasticity-mediated crosstalk
effect}
\author{Nadiv Dharan}
\affiliation{Department of Biomedical Engineering, Ben Gurion University,
Be'er Sheva 84105, Israel}
\author{Oded Farago}
\affiliation{Department of Biomedical Engineering, Ben Gurion University,
Be'er Sheva 84105, Israel}
\affiliation{Ilse Katz Institute for Nanoscale Science and Technology, Ben Gurion University of the Negev, Be'er
Sheva 84105, Israel}

\begin{abstract}
Cooperative action of molecular motors is essential for many cellular
processes.  One possible regulator of motor coordination is the
elasticity-mediated crosstalk (EMC) coupling between myosin II motors
whose origin is the tensile stress that they collectively generate in
actin filaments. Here, we use a statistical mechanical analysis to
investigate the influence of the EMC effect on the sarcomere - the
basic contractile unit of skeletal muscles. We demonstrate that the
EMC effect leads to an increase in the attachment probability of
motors located near the end of the sarcomere while, simultaneously,
decreasing the attachment probability of the motors in the central
part. Such a polarized attachment probability would impair the motors
ability to cooperate efficiently.  Interestingly, this undesired
phenomenon becomes significant only when the system size exceeds that
of the sarcomere in skeletal muscles, which provides an explanation
for the remarkable lack of sarcomere variability in
vertebrates. Another phenomenon that we investigate is the recently
observed increase in the duty ratio of the motors with the tension in
muscle. We reveal that the celebrated Hill's equation for muscle
contraction is very closely related to this observation.
\end{abstract}

\maketitle

\section{Introduction}

Skeletal striated muscles are tissues that contract in order to
produce the movement of the body. The fundamental contractile unit of
the muscle is the {\em sarcomere}\/ which is composed of two types of
filaments, actin and myosin, in an arrangement that allows them to
slide past each other \cite{keener}. Upon nervous stimulation, ${\rm
Ca}^{2+}$ ions flow into the muscle cell and expose binding sites
located on the actin filament \cite{alberts:2002}. The myosin thick
filament consists of myosin II motor proteins with a lever arm
structure that bind to the binding sites on the actin thin filament
and, via ATP hydrolysis, change its conformation, resulting in a
``power stroke'' that propels the myosin filament on top of the actin
filament \cite{cooke}. Comparison of skeletal muscle cells in
different vertebrates reveals that the lengths of their sarcomeres are
almost identical. The length of the thick filament is usually found to
be close to 1.6 $\mu$m, while the length of the thin filament is
typically in the $0.95-1.25$ $\mu$m range \cite{burkholder}.  This
fact is striking considering the different tasks that different
muscles perform in different species. Here, we argue that this
remarkable feature of muscles is closely related to the ability of the
myosin II motors to work in cooperation, which may be jeopardized by
the elasticity-mediated crosstalk (EMC) effect arising from the
compliance of the actin filament \cite{farago}. During muscle
contraction, the application of opposite forces (the motor forces
vs.~the external load) on the actin filaments causes large force
fluctuations. This leads to an increase in the elastic energy stored
in the filaments that can be lowered if the the duty ratios of the
myosin II motors are changed.  Our analysis shows that, in contracting
muscles, the EMC effect causes the attachment probability of motors to
become non-uniform (spatially dependent). This feature negatively
affect muscle performance since it hampers the ability of the motors
to cooperate efficiently. Interestingly this undesired phenomenon
becomes significant only when the system size exceeds that of the
sarcomere, which provides a plausible explanation for the similarity
of the sarcomeres in muscle cells across vertebrates.

Our current understanding of the mechanics of muscle contraction is
very much influenced by two classical works. The first one is
A.~V.~Hill's work (1938) \cite{hill}, in which the muscle was
represented through a combination of elastic, contractile, and
resistive (viscous) elements. Hill postulated that in overcoming the
viscous resistance, the contracting muscle does work and produces
heat. Through a general notion of energy balance and some empirical
relations between the rate of heat production during muscle
contraction and the contraction velocity, Hill derived his famous
equation:
\begin{equation}
(P+a)(v+b)=(P_0+a)b ,
\label{eq:hill}
\end{equation}
where $P$ is the load opposing the contraction, $v$ is the contraction
velocity, $P_0$ is the isometric load (i.e., the load for $v=0$) and
$a$ and $b$ are constants. Although generally considered as a
phenomenological force-velocity relationship rather than a
thermo-mechanical expression, Hill's equation has drawn much attention
because of its simplicity and the agreement it shows with experimental
measurements \cite{ieee}.

The second seminal work is Huxley's crossbridge theory from 1957 that
provides a molecular-level interpretation for muscle contraction
\cite{huxley:57}. Within the model, the myosin motor heads interact
with specific binding sites along the actin filament to form elastic
crossbridges. When a motor is attached to a binding site, the
crossbridge stretches and force is applied on the actin filament,
resulting in the relative movement of the actin (thin) and myosin
(thick) filaments past each other. The attachment and detachment of
motor heads to and from the actin filament are governed by ``on'' and
``off'' rate functions that regulate the fraction of crossbridges
(i.e., attached motors), and which depend on the stretching energy of
the crossbridges. The ``on'' and ``off'' rates were chosen by Huxley
to obtain a good fit with Hill's experimental data for the
force-velocity relationship [Eq.~(\ref{eq:hill})].

Huxley's work, together with the development of improved methods for
experimental determination of the sarcomere's micro-structure
\cite{huxley:63}, as well as new biochemical measurements of ATP
activity \cite{kendrick}, have provided a fruitful field for further
investigations. In 1971, a widely accepted four-state scheme for the
mechanochemical cycle of myosin II has been introduced by Lymn and
Taylor \cite{lymn}.  More recently Duke \cite{duke}, and later others
\cite{lan,kitamura, walcott,mansson,marcucci,chen}, have come up with
models that integrated the Lymn-Taylor scheme into the crossbridge
model. Duke's model (like Huxley's) treats the motor heads as elastic
elements with strain-dependent on and off rates. The model assumes
that the viscous friction forces can be neglected in the equation of
motion of the contracting muscle. The latter assumption remains
controversial, and several alternative models incorporating viscous
effects have been also proposed for the contractile process
\cite{worthingtonA,landesberg}. Both classes of models have been
successful in producing Hill's force-velocity relationship.

Another major success of the theoretical models is their ability to
show that the fraction of working (force-producing) motors, $r$,
increases with the load $P$. This feature has been recently observed
by Lombardi and co-workers \cite{lombardi} who showed, by using X-ray
scattering and mechanical measurements on tibialis anterior muscles of
frogs, that $r$ increases from roughly $r=0.05$ for $P=0$ to $r=0.3$
for the maximal isometric load. Related patterns of collective
behavior also emerge in models for motor protein motility assays
\cite{badoual,julicher}. Specifically, several ratchet models have
shown that when two groups of antagonistic motors are engaged in a
``tug-of-war'' competition, their detachment rates may be considerably
varied \cite{lipowsky,kafri,gilboa}. Another setup, closely related to
muscle contractility, is a single class of motors that work against
the force produced by an optical trap \cite{joanny}. One of the
factors that significantly alters the detachment rates is the
elasticity-mediated crosstalk (EMC), whose effect on muscle
contractility is studied here using a simple spring-bead
model.

\section{Hill's equation} 

Before presenting our model and its results, we first wish to examine
more closely the experimental results of ref.~\cite{lombardi}. To this
end, it is useful to define the dimensionless variables $0\leq x\equiv
P/P_0\leq 1$, and $0\leq y\equiv v/v_{\rm max}\leq 1$, where $v_{\rm
max}$ is the maximum contraction velocity at $P=0$. When expressed in
terms of these variables, Hill's equation (\ref{eq:hill}) takes the
dimensionless form:
\begin{equation}
y=\frac{1-x}{1+cx},
\label{eq:hill2}
\end{equation}
where $c$ is a constant. Notice that eq.~(\ref{eq:hill2}) satisfies
both the relation that $y=1$ for $x=0$ (load-free contraction) and
$y=0$ for $x=1$ (isometric contraction).  The other important notion
in relation to muscle contraction is the observed (see Fig.~3(D) in
ref.~\cite{lombardi}) increase in $r$ with $P$, which is well
approximated by the linear relationship
\begin{equation}
r=r_0+(r_1-r_0)x,
\label{eq:linear_r}
\end{equation}
where $r_0$ and $r_1$ denote the attachment probability for $x=0$ and
$x=1$, respectively. To match the experimental data, we set $r_0=0.05$
and $r_1\simeq 0.3$.  Since the sarcomere contracts at a constant
velocity, then Newton's first law of motion implies that the forces
generated by the motors are balanced by the external load $P$ and the
friction forces in system. The latter originate from two sources: the
surrounding medium and the crossbridges. The balance of forces reads
\begin{equation}
r(f_m-\lambda_{m}v)=\lambda v+\frac{P}{N}.
\label{eq:newton}
\end{equation}
The expression in brackets on the left hand side of
eq.~(\ref{eq:newton}) can be identified as the force per motor
\begin{equation}
f_0(v)=f_m-\lambda_{m}v.
\label{eq:linear_f}
\end{equation}
It includes a positive ``active force'', $f_m$, and a negative
``motor-friction force'' characterized by the motor friction
coefficient $\lambda_m$ \cite{note}. This linear force-velocity
relationship is consistent with the experimental results shown in
Fig.~4(B) of ref.~\cite{lombardi}. On the right hand side of
eq.~(\ref{eq:newton}) we have the counter external load, $P/N$, and
the friction force caused by the surrounding medium, $\lambda v$
($\lambda \neq \lambda_m$), both normalized per motor.

It is important to emphasize that eq.~(\ref{eq:linear_r}) is
empirical, and we do not intend to discuss its physical basis. Here,
we simply want to demonstrate that Hill's dimensionless equation can
be derived from eqs.~(\ref{eq:linear_r}) and (\ref{eq:newton}) without
any further assumptions. Explicitly, upon substitution of
eq.~(\ref{eq:linear_r}) in eq.~(\ref{eq:newton}) and rearrangement of
the resulting equation, one arrives to the following expression for
the contraction velocity
\begin{equation}
v=\frac{f_mr_0(1-x)}{\lambda+\lambda_m[r_0+(r_1-r_0)x]}.
\label{eq:v}
\end{equation}
Also, for $P=0$, eq.~(\ref{eq:newton}) takes the form
\begin{equation}
\frac{f_mr_0}{v_{\rm max}}=\lambda+\lambda_mr_0.
\label{eq:newton_0}
\end{equation}
Dividing eq.~(\ref{eq:v}) by $v_{max}$, and using
eq.~(\ref{eq:newton_0}) as well as the relation $P/N=xP_0/N=xf_mr_1$,
one arrives to eq.~(\ref{eq:hill2}) with
\begin{equation}
c=\frac{\lambda_m (r_1-r_0)}{\lambda+\lambda_mr_0}.
\label{eq:c}
\end{equation}
We further note that for $y=1$ ($v=v_{\rm max}$), the following
expression can be derived for $f_0$ if eq.~(\ref{eq:newton_0}) is used
in eq.~(\ref{eq:linear_f})
\begin{equation} 
f_0|_{y=1}=f_m\frac{\lambda}{\lambda+r_0\lambda_m}.
\label{eq:f_vmax}
\end{equation}
The experimental data \cite{lombardi} gives $f_m=6$pN,
$f_0|_{y=1}=4$pN, and $r_0=0.05$ which, upon substitution in
eq.~(\ref{eq:f_vmax}), yields $\lambda_m=10\lambda$. When this last
result, together with the values for $r_0$ and $r_1$, are used in
eq.~(\ref{eq:c}), one arrives to $c=5/3$, which is within the range
$1.2<c<4$ where the constant $c$ is typically found for skeletal
muscles \cite{fung}. Notice that in order to match the experimental
data, we included viscous terms in our derivation, which (as
noted above) is a matter of controversy. Nevertheless, it is important
to recognize that eq.~(\ref{eq:c}) has a well defined limit when both
$ \lambda$ and $\lambda_m$ vanish, provided that $\lambda/\lambda_m$
is finite.

\begin{figure}[t]
\begin{center}
\scalebox{0.375}{\centering \includegraphics{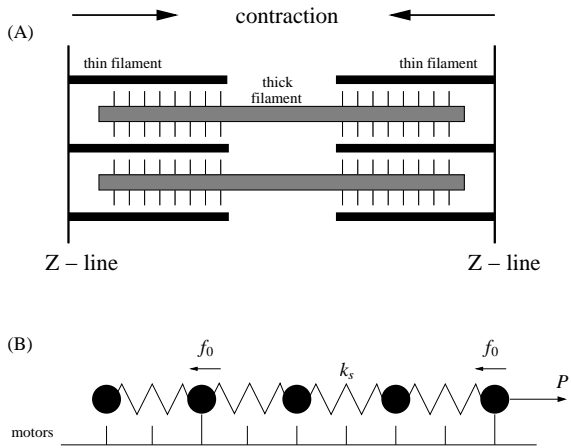}}
\end{center}
\vspace{-0.5cm}
\caption{(A) Schematics of the sarcomere, consisting of an array of
parallel actin thin filaments surrounded by thick filaments of myosin
motors. Adjacent sarcomeres are connected end-to-end at the {\em
Z-line}. (B) In our model, the actin filament is represented as a
chain of nodes, connected by identical springs with spring constant
$k_s$. Each node is either connected to a single myosin II motor - in
which case it experiences a force of magnitude $f_0$, or disconnected
- in which case it experiences no force. The motor forces are
countered by the external force $P$ which acts at the end of the
chain.}
\label{fig:sarcomere}
\end{figure}

\section{Model} 

We now proceed to discuss the relevance of the EMC effect to skeletal
muscle contraction. Fig.~\ref{fig:sarcomere}(A) shows a schematics of
the sarcomere structure, consisting of an array of parallel actin thin
filaments surrounded by thick filaments of myosin motors. Adjacent
sarcomeres are connected end-to-end to form {\em myofibrils}. The ends
of the actin filaments are anchored at the {\em Z-line}\/, which
transmits the external load to the actin filaments. Our model of the
sarcomere is depicted in Fig.~\ref{fig:sarcomere}(B). The elastic
actin filament is modeled as a chain of $N$ monomers connected by
$N-1$ elastic springs with spring constant $k_s$. We assume that a
motor exerts a force of magnitude $f_0$ on the actin filament in the
attached state, and no force in the detached state. The other model
parameter is the attachment probability of the motors, $r$. As
discussed above, both $f_0$ and $r$ vary with $v$. Our model neglects
spatial (motor-to-motor) and temporal variations of $r$ and $f_0$ {\em
at a given}\/ $v$, so these two quantities represent the typical
attachment probability and motor force, respectively. As we will
demonstrate below, it is the EMC effect that leads to spatial
variations in the {\em effective}\/ (mean) attachment probability of
the motors, which we will denote by $\langle r\rangle$ (to be
distinguished from the uniform ``bare'' attachment probability~$r$).

The motor forces work against the opposite external force which is
applied on the $N$-th last monomer. The tug-of-war competition between
the motor forces and the external force stretches the actin
filament. Denoting by $E^{\rm el}$ the elastic energy of the filament,
the statistical weight of a configuration with $n$ connected motors is
given by $w=r^n(1-r)^{N-n}\exp(-E^{\rm el}/k_BT)$, where $k_BT$ is the
thermal energy. We treat the elastic energy as an equilibrium degree
of freedom (of a system which is inherently out-of-equilibrium)
because the mechanical response of the filament to the attachment or
detachment of motors is extremely rapid and occurs on time scales
which are far shorter than the typical attachment time of the
motors. The elastic energy $E^{\rm el}$ is calculated as follows
\cite{dharan}. We denote by $f_i$ the force applied on the $i$-th
monomer, where $f_i=0$ or $f_i=f_0$ for $i=1,\ldots,(N-1)$, and,
$f_i=-P$ or $f_i=-P+f_0$ for $i=N$. The total elastic energy is the
sum of spring energies, $E^{\rm el}=\sum_{i=1}^{N-1}F^2_i/2k_s$, where
$F_i$ is the force applied on the $i$-th spring. The forces $F_i$ are
calculated as follows: We first calculate the mean force $\bar
{f}=\left(\sum_{i=1}^N f_i\right)/N$, and define the excess forces
acting on the nodes: $f_i^*=f_i-\bar{f}$. The force on the $i$-th
spring is then obtained by summing the excess forces applied on all
the monomers located on one side of the spring:
$F_i=-\sum_{l=1}^{i}f^*_l=\sum_{l=i+1}^Nf^*_l$.

The model is studied by using Monte Carlo simulations with trial moves
that attempt to change the state (connected or disconnected) of a
randomly chosen motors. For each move attempt, the elastic energy
$E^{\rm el}$ of the chain is recalculated, and the move is accepted or
rejected according to the conventional Metropolis criterion with the
statistical weights given by $w$.

\begin{figure*}[t]
\begin{center}
\scalebox{0.75}{\centering \includegraphics{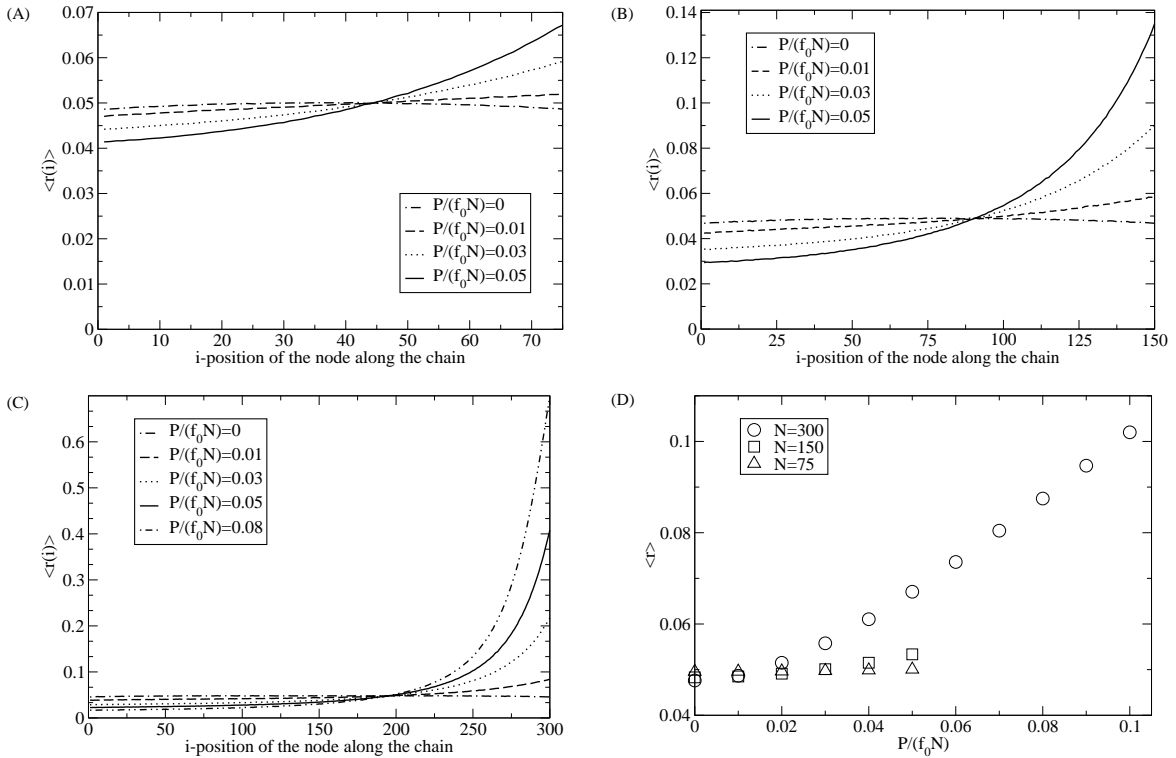}}
\end{center}
\vspace{-0.5cm}
\caption{The attachment probability $\langle r(i)\rangle$ to the
$i$-th chain node, calculated for chains consisting of (A) $N=75$, (B)
$N=150$, and (C) $N=300$ nodes. For each $N$, the attachment
probability is plotted for various values of the load $P$. (D) The
effective attachment probability $\langle r\rangle$ as a function of
the dimensionless load per motor $P/(f_0N)$ for $N=75$ (triangles),
$N=150$ (squares), and $N=300$ (circles).}
\label{fig:r_i}
\end{figure*}

\section{Results} 

In half sarcomere, each thick filament has about 150 motor heads
\cite{alberts:2002}. To simulate muscles operating under conditions of
optimal force generation, we assume that there is a full overlap
between the thick and thin filaments \cite{keener}, and consider a
chain with $N=150/2=75$ monomers. (The division by $2$ is due to the
1:2 ratio between thin and thick filaments) We also simulate larger
systems of $N=150$ and $N=300$ nodes and fix the model parameters to
$r=0.05$, $f_0=f_m=6$ pN, and $k_s\simeq 4.5$ N/m (see discussion on
how these values were set in \cite{dharan}). Our simulation results
for the mean attachment probability, $\langle r(i)\rangle$, as a
function of $i$ ($1\leq i\leq N$), the position of the monomer along
the chain, are depicted in Fig.~\ref{fig:r_i} for $N=75$ (A), $N=150$
(B), and $N=300$ (C). The simulations reveal that due to the EMC
effect, the attachment probability becomes a monotonically increasing
function of $i$. The origin of this feature is the fact that the
springs are not equally stretched, as can be inferred from the above
derivation of the elastic energy. Generally speaking, attachment of a
motor to a certain node $i$ leads to a reduction in the energy of the
springs with $j<i$. For each $N$, there is a single node ($i=i^*$)
where the attachment probability, $\langle r(i^*)\rangle$, is
independent of $P$ and takes a value which is very close to the bare
attachment probability $r$. The difference between the attachment
probabilities at both ends of the chain (i.e., for $i=1$ and $i=N$)
increases with both $N$ and $P$. For $N=75$, the variation in $\langle
r(i)\rangle$ is quite small, becoming meaningful only at near-stall
forces $P/(f_0N)\simeq 0.05$. In contrast, for $N=150$, the variations
in $\langle r(i)\rangle$ are significant and may be as large as
$\langle r(N)\rangle/\langle r(1)\rangle \gtrsim 4$. The mean
attachment probability, $\langle r\rangle\equiv [\sum_{i=1}^N
r(i)]/N$, is plotted in Fig.~\ref{fig:r_i}(D) as a function of
$P$. For both $N=75$ and $N=150$ and for all values of $P$, we find
that $\langle r\rangle\simeq r$, which shows that the decrease in
$\langle r(i)\rangle$ for $i<i^*$ is almost offset by the increase in
$\langle r(i)\rangle$ for $i>i^*$. This is not the case for $N=300$,
where $\langle r\rangle$ exhibits a steady increase with $P$. The
increase in $\langle r\rangle$ is due to the fact that at large forces
(notice that the applied loads are proportional to $N$), the
attachment probability $\langle r(i)\rangle$ becomes a rapidly
increasing function of $i$. The implication of the rise in $\langle
r\rangle$ is that the stall force {\em per motor}\/ increases to
$\simeq 0.1 f_0$ (from $\simeq 0.05 f_0$ for $N=75$ and $N=150$).

%\begin{widetext}

%\end{widetext}
\begin{figure}[t]
\begin{center}
\vspace{0.3cm}
\scalebox{0.45}{\centering \includegraphics{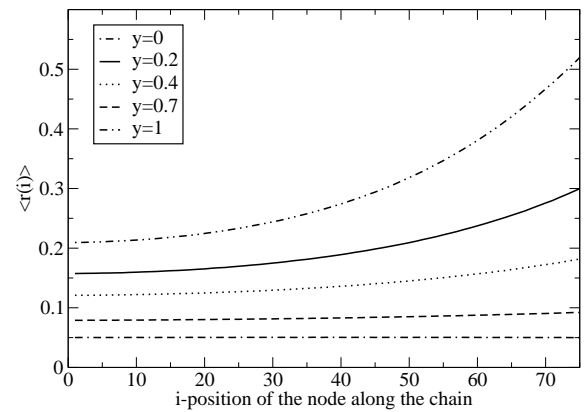}}
\end{center}
\vspace{-0.5cm}
\caption{The attachment probability $\langle r(i)\rangle$ to the
$i$-th chain node, calculated for different values of the rescaled
velocity $y$.}
\label{fig:fig3}
\end{figure}

The above results seem to indicate that for $N>75$, the variations in
$\langle r(i)\rangle$ may be sufficient to disrupt muscle
performance. However, one needs to recall that the simulation results
depicted in Fig.~\ref{fig:r_i} correspond to fixed values of $r$ and
$f_0$, while in reality the values of these quantities vary with the
shortening velocity $v$. Therefore, we also performed simulations
where for each value of $P$, the appropriate value of $v$ is chosen
and, accordingly, the values of $r$ and $f_0$ are set. This has been
done by using eqs.~(\ref{eq:hill2}), (\ref{eq:linear_r}), and
(\ref{eq:linear_f}), as outlined in Appendix \ref{append}. Our results
for the sarcomere simulations are summarized in Fig.~\ref{fig:fig3}
which shows the position-dependent attachment probability, $\langle
r(i)\rangle$, calculated for a chain of size $N=75$ and for different
loads. Similarly to the results plotted in the above
Fig.~\ref{fig:r_i}(A), the results here indicate that $\langle
r(i)\rangle$ remains fairly uniform for $1\leq i\leq N=75$. This
observation is consistent with the motor mechanochemistry, where at
high and medium velocities $v$ the detachment of motors occurs after
the completion of the power stroke. In the mechnochemical picture, the
detachment rate can be estimated by $\tau_{\rm det}^{-1}\sim v/\delta$
where $\delta$ is the size of the power stroke ($\delta \sim 6$ nm
\cite{lombardi}), and this rate must be the same for all the
motors. Highly polarized detachment probability hampers the motors
ability to cooperate and, thus, constitutes an undesired
effect. Notice that Fig.~\ref{fig:fig3} features a small increase in
the attachment rate near the Z-line (for $y>0.2$). This, in fact,
represents a ``positive'' effect, as it enables the muscle to sustain
a larger load. We stress that the discussion here is relevant only for
high and medium velocities where the motors attach, execute the power
stroke, and then unbind. At very low velocities ($y\simeq 0$) close to
the isometric load ($x\simeq 1$), the motors may detach and reattach
multiple times before the power stroke is completed
\cite{lombardi2}. The number of attachments per power stroke need not
be the same for all the motors and, therefore, variations in the
attachment probabilities (which are indeed observed in
Fig.~\ref{fig:fig3} at small values of $y$) can be tolerated in this
limit without negative consequences for muscle contractility.

\section{Summary} 

We found that the EMC effect leads to an increase in the attachment
probability (duty ratio) of motors located near the Z-line while,
simultaneously, decreasing the attachment probability of the motors in
the central part of the sarcomere (M-line). The resulting variations
in the attachment probabilities of the motors pose a serious problem
for muscle performance because the detachment rate of the motors is
primarily determined by the size of the power stroke and the muscle
shortening velocity and, therefore, should be fairly uniform. We used
a simple bead-spring model to demonstrate that this undesired effect
can be almost neglected for half-sarcomeres consisting of less than
approximately $2N=150$ motors, which is precisely the ``universal''
size of half a sarcomere. Recent advances in single molecule
techniques \cite{schmidt} open the opportunity for testing our
theoretical predictions concerning the polarization of the attachment
probability. Specifically, optical tweezers provide the mean to apply
pico-newton forces against the gliding motion of the thin filament,
while the duty ratio may be directly measured using high resolution
atomic force microscopy (AFM). 

We thank Sefi Givli for numerous insightful discussions.

\appendix
\section{Details of the sarcomere simulations}
\label{append}

In the simulations depicted in Fig.~\ref{fig:fig3}, the values of $x$
(rescaled load), $y$ (rescaled velocity), $f_0$ (motor force), and $r$
(duty ratio) are different for each curve. Their values are set by
using eqs.~(\ref{eq:hill2}), (\ref{eq:linear_r}), and
(\ref{eq:linear_f}), with system parameters that are chosen to match
the experimental results of ref.~\cite{lombardi}. For a given value of
$y$, the value of $x$ is determined from Hill's equation
(\ref{eq:hill2})
\begin{equation} 
y=\frac{1-x}{1+(5/3)x},
\end{equation}
while the value of $f_0$ is set according to eq.~(\ref{eq:linear_f})
\begin{equation}
f_0=6(1-\frac{y}{3})\ {\rm [pN]}.
\label{eq:f0}
\end{equation} 
The determination of $r$ is slightly more complicated. The
complication arises due to the fact that the bare attachment
probability $r$ appearing in eqs.~(\ref{eq:linear_r}) and
(\ref{eq:newton}) should be replaced with the effective attachment
probability, $\langle r\rangle$, which is the experimentally measured
quantity and which is not known a-priori. Therefore, for each value of
$y$ we simulated different values of $r$ and, using linear
interpolation, determined the value of $r$ that yields the desired
$\langle r\rangle$. As can be inferred from Fig.~\ref{fig:fig3}, the
difference between $r$ and $\langle r\rangle$ is quite small for high
and medium velocities.

Another important comment is the following: If the excess forces used
in the calculation of the elastic energy are expressed in units of the
motor force $f_0$, $\tilde{F}_i\equiv F_i/f_0$, then the Boltzmann
weight of a configuration with $n$ connected motors can be written as
\begin{equation}
w=r^n(1-r)^{N-n}\exp\left(-\beta^*\sum_{i=1}^{N-1}\tilde{F}_i^2\right),
\end{equation}
where $\beta^*=f_0^2/2k_sk_BT$. In the simulations presented in
Fig.~\ref{fig:r_i}, the motor force is taken as $f_0=f_m=6$~pN, while
the spring constant $k_s=4.5$ N/m. This means that at physiological
temperature of $T=310K$, $\beta^*\simeq 10^{-3}$, which is the value
of $\beta^*$ used in the simulations depicted in
Fig.~\ref{fig:r_i}. In Fig.~\ref{fig:fig3}, we use the above
Eq.~\ref{eq:f0} to evaluate $f_0$. This means that for each curve in
Fig.~\ref{fig:fig3}, the dimensionless parameter $\beta^*$ has been
redefined to $\beta^*=10^{-3}(1-y/3)^2$.

%----------------------------------------------------------- 
%References
%----------------------------------------------------------- 


\begin{thebibliography}{99}
\bibitem{note} On the microscopic time scales of the power stroke, the
velocity of the motor head changes and the motor force fluctuates
between positive and negative values. In contrast, Hill's equation
applies to a coarse-grained picture where the thick and thin filaments
are treated as two objects sliding at relative
velocity $v$. Therefore, the motor force expressed in eq.~(5)
should be considered as an effective force that each attached motor
generates between these two objects.
\end{thebibliography}
\end{document}